\documentclass[aip,jcp,preprint,letterpaper,onecolumn]{revtex4-1}

\usepackage{booktabs}
\usepackage{amssymb}
\usepackage{amsmath}
\usepackage[usenames,dvipsnames]{xcolor}
\usepackage{graphicx}
\usepackage{float}
\usepackage[normalem]{ulem}
\usepackage{subcaption}
\usepackage{epstopdf}

\usepackage{natbib}
\newcommand{\bra}[1]{\left\langle #1 \right|}
\newcommand{\ket}[1]{\left|#1\right\rangle}

\newcommand{\dens}[2]{\ket{#1}\bra{#2}}
\newcommand{\abs}[1]{\left|#1\right|}

\usepackage{xcolor}

\usepackage{multirow}
\usepackage{bm}

\begin{document}

\title{Too big, too small or just right? A benchmark assessment of density functional theory for predicting the spatial extent of the electron density of small chemical systems.}

\author{Diptarka Hait}
\email{diptarka@berkeley.edu}
\altaffiliation{These authors contributed equally to this work.}
\affiliation
{{Kenneth S. Pitzer Center for Theoretical Chemistry, Department of Chemistry, University of California, Berkeley, California 94720, USA}}
\affiliation{Chemical Sciences Division, Lawrence Berkeley National Laboratory, Berkeley, California 94720, USA}

\author{Yu Hsuan Liang}
\altaffiliation{These authors contributed equally to this work.}
\affiliation
{{Kenneth S. Pitzer Center for Theoretical Chemistry, Department of Chemistry, University of California, Berkeley, California 94720, USA}}

\author{Martin Head-Gordon}
\email{mhg@cchem.berkeley.edu}
\affiliation
{{Kenneth S. Pitzer Center for Theoretical Chemistry, Department of Chemistry, University of California, Berkeley, California 94720, USA}}
\affiliation{Chemical Sciences Division, Lawrence Berkeley National Laboratory, Berkeley, California 94720, USA}

	\begin{abstract}
     Multipole moments are the first-order responses of the energy to spatial derivatives of the electric field strength. The quality of density functional theory (DFT) prediction of molecular multipole moments thus characterizes errors in modeling the electron density itself, as well as the performance in describing molecules interacting with external electric fields. 
     However, only the lowest non-zero moment is translationally invariant, making the higher-order moments origin-dependent. Therefore, instead of using the $3 \times 3$ quadrupole moment matrix, we utilize the translationally invariant $3 \times 3$ matrix of second \textit{cumulants} (or spatial variances) of the electron density as the quantity of interest (denoted by $\bm{\mathcal{K}}$). The principal components of $\bm{\mathcal{K}}$ are the square of the spatial extent of the electron density along each axis. A benchmark dataset of the principal components of $\bm{\mathcal{K}}$ for \textcolor{black}{100} small molecules at the coupled cluster singles and doubles with perturbative triples (CCSD(T)) at the complete basis set (CBS) limit is developed, resulting in \textcolor{black}{213} independent $\bm{\mathcal{K}}$ components.
     The performance of 47 popular and recent density functionals is assessed against this \textcolor{black}{Var213} dataset. Several functionals, especially double hybrids, and also SCAN and SCAN0 yield reliable second cumulants, although some modern, empirically parameterized functionals yield more disappointing performance. The H and Be atoms in particular are challenging for nearly all methods, indicating that future functional development could benefit from inclusion of their density information in training or testing protocols. 
	\end{abstract}
	\maketitle
\section{Introduction}

Quantum chemistry methods have seen increasingly widespread use over the last few decades, and now permit sufficiently accurate computation of energies ranging in scale from total atomization energies\cite{karton2011w4} to extremely weak noncovalent interactions\cite{rezac2013describing}. In particular, density functional theory\cite{becke2014perspective,jones2015density,mardirossian2017thirty} (DFT) is ubiquitously employed to study medium to large systems as it provides an acceptable balance between accuracy and computational cost. Although exact in theory\cite{hohenberg1964inhomogeneous,levy1979universal}, practical usage of DFT almost always entails use of computationally tractable density functional approximations (DFAs) within the Kohn-Sham (KS) formalism\cite{kohn1965self}. It is often remarked that DFAs are not really systematically improvable the way the exact coupled cluster (CC) hierarchy is, but the best-performing DFAs on a given rung of Jacob's ladder\cite{perdew2001jacob} statistically improve upon lower rungs\cite{mardirossian2017thirty,goerigk2017look,mardirossian2018survival,najibi2018nonlocal}, indicating that greater complexity \textit{could} (but is not guaranteed to) lead to better accuracy. Caution is still warranted while applying KS-DFT to systems with substantial levels of delocalization error\cite{perdew1982density,mori2006many,hait2018delocalization,lonsdale2020one} or multireference character\cite{cohen2008fractionalspin,cohen2008insights,hait2019wellbehaved}, but DFAs are quite effective overall at predicting ground state relative energies associated with main-group chemistry\cite{mardirossian2017thirty,goerigk2017look}, and even organometallic chemistry.\cite{dohm2018mori41,chan2019tmc151}  Indeed, modern DFAs often surpass the accuracy of single-reference wave function approaches like second-order M{\o}ller-Plesset perturbation theory (MP2) or CC singles and doubles (CCSD) for such problems\cite{mardirossian2018survival,hait2018accurate,rettig2020third}.

This is perhaps unsurprising, as DFA development has historically emphasized improved prediction of such chemically relevant energies, out of the belief that this leads to better approximations to the exact functional. However, the exact functional ought to map the exact density to the exact energy, and it is entirely possible for a DFA to obtain a reasonable energy from a relatively inaccurate density via hidden cancellation of errors. It has indeed been suggested that modern DFAs predict worse densities than older, less parameterized models\cite{medvedev2017density}-leading to considerable discussion\cite{hammes2017conundrum,kepp2017comment,medvedev2017response,graziano2017quantum,korth2017density,brorsen2017accuracy,wang2017well,ranasinghe2017note,mayer2017conceptual,gould2017makes,hait2018accurate} in the scientific community about density predictions from DFT. It is certainly plausible that greater complexity in the functional form creates additional avenues for incorrect behavior (as can be seen in wave function theory as well\cite{leininger2000mo,hait2019levels}), even without accounting for overfitting to empirical data. 

Separate from the debate about ``true paths'' for DFA development,\cite{medvedev2017density} we note that densities are important for a very practical reason. The electron density controls the response of the electronic energy to external electric fields, such as those that might be encountered in condensed phases or in spectroscopic simulations. Density errors are thus related to energy errors under a different one-particle potential (that DFT should still be formally capable of addressing\cite{hohenberg1964inhomogeneous}). Application of DFT beyond ground state gas-phase chemical physics is thus reliant upon DFT yielding reasonable densities, or at least a characterization of when and how various DFAs fail. This has led to investigations on the performance of DFT for predicting dipole moments\cite{hickey2014benchmarking,verma2017can,hait2018accurate,johnson2019effect,grotjahn2020evaluation,hait2018communication} and polarizabilities\cite{hickey2014benchmarking,thakkar2015well,hait2018accuratepolar,grotjahn2020evaluation,withanage2019self,hait2019beyond}, as they represent the linear and quadratic responses, respectively, of the energy to a spatially uniform electric field. 

The natural next step would be to assess the linear response of the energy to a (spatially) constant electric field gradient, which is equivalent to the molecular quadrupole moment $\mathbf{Q}$. Indeed, there have been a number of assessments of the quality of various density functionals for evaluating molecular quadrupole moments in the past\cite{cohen1999,deproft2000,hofinger2002,matta2010}, albeit before the development of many modern functionals. At the same time, efforts have been made to converge high-quality coupled cluster theory calculations of quadrupole moments with respect to basis set,\cite{halkier1997props,zhang2016props,bokhan2016props} setting the stage for benchmark assessments. $\mathbf{Q}$ is however not translationally invariant for polar systems, making it a somewhat unsuitable metric. We thus decided to investigate the performance of DFT in predicting the second spatial cumulant of the electron density ($\bm{\mathcal{K}}$) instead, which is a translationally invariant quantity connected to $\mathbf{Q}$  and the dipole moment $\bm{\mu}$. $\bm{\mathcal{K}}$ thus relates the density to the response of the energy to electric fields that linearly vary with spatial coordinates. Mathematically, $\bm{\mathcal{K}}$ is also the variance of the electron density. To the best of our knowledge, there have been no previous studies that explicitly focus on prediction of $\bm{\mathcal{K}}$ by DFT or \textcolor{black}{MP/CC} methods. \textcolor{black}{However, a related quantity had previously been studied with Hartree-Fock (HF)\cite{hollett2006quantum} and found to be useful in predicting steric effects of substituents}.

\section{Second spatial cumulant of electron density}
\subsection{Definition}
Let us consider a system with electron density $\rho(\mathbf{r})$ and total number of electrons $N=\displaystyle\int \rho(\mathbf{r}) d\mathbf{r}$. The spatial probability density $p(\mathbf{r})$ for a single electron is:
\begin{align}
    p(\mathbf{r}) = \dfrac{\rho(\mathbf{r})}{N}
\end{align}
An alternative definition of $p(\mathbf{r})$ in terms of the wave function $\ket{\Psi}$ is:
\begin{align}
    p(\mathbf{r}) & = \bra{\mathbf{r}}\left(\underset{N-1}{\mathbf{Tr}}\left[\dens{\Psi}{\Psi}\right]\right) \ket{\mathbf{r}} = \dfrac{\Gamma_1(\mathbf{r},\mathbf{r})}{N}
\end{align}
which corresponds to the scaled diagonal elements of the one particle density operator, $\Gamma_1(\mathbf{r},\mathbf{r'})$ that results from tracing out $N-1$ degrees of freedom from the full density operator $\Gamma_N=\dens{\Psi}{\Psi}$. For a single Slater determinant, this is equivalent to averaging the square of each occupied spin orbital. 

The first and second spatial moments of the electron density are consequently:
\begin{align}
    \langle \mathbf{r}\rangle &= \displaystyle\int \mathbf{r} p(\mathbf{r})d\mathbf{r}\\
    \langle \mathbf{r}\mathbf{r^T}\rangle &= \displaystyle\int \left( \mathbf{r}\mathbf{r^T}\right) p(\mathbf{r})d\mathbf{r}
\end{align}

The second spatial cumulant of the electron density is thus:
\begin{align}
    \bm{\mathcal{K}}&=\langle \left(\delta \mathbf{r}\right)\left(\delta \mathbf{r}\right)^\mathbf{T}\rangle= \langle \mathbf{r}\mathbf{r^T}\rangle- \langle \mathbf{r}\rangle \langle \mathbf{r}\rangle^\mathbf{T}
\end{align}
which is equivalent to the spatial (co)variance of the probability distribution $p(\mathbf{r}$). The individual components are thus of the form:
\begin{align}
    \mathcal{K}_{xx} &=\langle x^2 \rangle-\langle x \rangle^2=\displaystyle\int x^2 p(\mathbf{r})d\mathbf{r}-\left(\displaystyle\int x p(\mathbf{r})d\mathbf{r}\right)^2\\
    \mathcal{K}_{xy} &=\langle xy \rangle-\langle x \rangle\langle y \rangle=\displaystyle\int xy p(\mathbf{r})d\mathbf{r}-\left(\displaystyle\int x p(\mathbf{r})d\mathbf{r}\right)\left(\displaystyle\int y p(\mathbf{r})d\mathbf{r}\right)
\end{align}
etc., in terms of the individual cartesian directions $x,y,z$. 
$\bm{\mathcal{K}}$ is thus translationally invariant, and represents the ``width" (or ``spread") of the electron density. 
We also note that the eigenvalues of $\bm{\mathcal{K}}$ are rotationally invariant. In addition, while we have only focused on the second-order cumulant in this work, higher-order cumulants of $p(r)$ can similarly be readily defined and utilized, for problems where they might be relevant. 

\subsection{Connection to Multipole Moments}
If the system has nuclear charges $\{Z_A\}$ at positions $\{\mathbf{R_A}\}$, the dipole moment $\bm{\mu}$ and (cartesian/non-traceless) quadrupole moment $\mathbf{Q}$ are:
\begin{align}
    \bm{\mu} & = \displaystyle\sum\limits_A Z_A\mathbf{R_A} -\displaystyle\int\mathbf{r} \rho(\mathbf{r})d\mathbf{r}=\displaystyle\sum\limits_A Z_A\mathbf{R_A} -N\langle \mathbf{r}\rangle \\
    \mathbf{Q} & = \displaystyle\sum\limits_A Z_A\left(\mathbf{R_A}\right)\left( \mathbf{R_A}\right)^\mathbf{T} -\displaystyle\int \left(\mathbf{r}\mathbf{r^T}\right) \rho(\mathbf{r})d\mathbf{r}=\displaystyle\sum\limits_A Z_A\left(\mathbf{R_A}\right)\left( \mathbf{R_A}\right)^\mathbf{T} -N\langle \mathbf{r}\mathbf{r^T}\rangle
\end{align}

Therefore the second cumulant can be expressed as:
\begin{align}
    \bm{\mathcal{K}}&=\dfrac{1}{N}\left(\displaystyle\sum\limits_A Z_A\left(\mathbf{R_A}\right)\left( \mathbf{R_A}\right)^\mathbf{T} - \mathbf{Q} \right)-\dfrac{1}{N^2}\left(\displaystyle\sum\limits_A Z_A\mathbf{R_A}- \bm{\mu} \right)\left(\displaystyle\sum\limits_A Z_A\mathbf{R_A} - \bm{\mu} \right)^\mathbf{T}
\end{align}



\subsection{Physical interpretation}
Let us begin with the case of atoms, and suppose that the single electron probability distribution for a (spherical) atom at $(0,0,0)$ is $p_1(\mathbf{r})$. Spherical symmetry indicates that:
\begin{align}
   \displaystyle\int  xp_1(\mathbf{r})d\mathbf{r}  &=
   \displaystyle\int   yp_1(\mathbf{r})d\mathbf{r} =
   \displaystyle\int   zp_1(\mathbf{r})d\mathbf{r} = 0\label{mom1}\\
   \displaystyle\int xyp_1(\mathbf{r})d\mathbf{r} &= 
   \displaystyle\int  yzp_1(\mathbf{r})d\mathbf{r} =
   \displaystyle\int  xzp_1(\mathbf{r})d\mathbf{r} = 0\label{offmom2}\\
   \displaystyle\int x^2p_1(\mathbf{r})d\mathbf{r} &= 
   \displaystyle\int  y^2p_1(\mathbf{r})d\mathbf{r} = 
   \displaystyle\int  z^2p_1(\mathbf{r})d\mathbf{r} = \eta^2 \label{atomsize}
\end{align}
This implies that  $\mathcal{K}_{xx}=\mathcal{K}_{yy}=\mathcal{K}_{zz}=\eta^2$. Therefore $\sqrt{3}\eta$ is an effective atomic radius associated with the spatial extent of the electron density, much like a covalent or \textcolor{black}{van} der Waals radius. In particular for the H atom, $\eta$ is the Bohr radius (1 a.u.=0.529 {\AA}).

For molecules without symmetry, $\bm{\mathcal{K}}$ is not generally a diagonal $3 \times 3$ matrix, and likewise $\mathcal{K}_{xx} \neq \mathcal{K}_{yy} \neq \mathcal{K}_{zz}$. $\bm{\mathcal{K}}$ can be diagonalized, to yield principal axes, and 3 principal components, very analogous to the molecular inertia tensor associated with the nuclei, $\mathbf{I} = \sum\limits_A m_A\left(\mathbf{R_A}\right)\left( \mathbf{R_A}\right)^\mathbf{T}$. The eigenvalues of $\bm{\mathcal{K}}$ thereby give the square of the spatial extent of the electron density along each principal axis. Systems where both a C$_n$ ($n>1$) axis of rotation and $\sigma_v$ plane(s) of symmetry are present (i.e. C$_{nv}$ or higher symmetry)  
have the principal axes be defined by molecular symmetry, and we shall later choose members of our data set based on this convenient simplification.


\subsection{Behavior vs system size}
The behavior of $\bm{\mathcal{K}}$ vs system size could be useful in characterizing the utility of this quantity. A simple case to consider is a 1D lattice of $M$ noninteracting He atoms, placed at $(0,0,0),(0,0,a),(0,0,2a)\ldots (0,0,(M-1)a)$. Let each atom have spatial extent $\eta$ as in Eq. \eqref{atomsize}.

Subsequently, $p(\mathbf{r})$ for the supersystem is:
\begin{align}
    p(\mathbf{r}) & = \dfrac{1}{M}\displaystyle\sum\limits_{m=0}^{M-1} p_1(\mathbf{r}-ma\hat{\mathbf{z}})
\end{align}
which is the average of the individual single electron probability distributions. 
Transverse to the lattice vector $a\hat{\mathbf{z}}$, we thus have:
\begin{align}
    \langle x^2 \rangle & = \dfrac{1}{M}\displaystyle\sum\limits_{m=0}^{M-1}\displaystyle\int x^2 p_1(\mathbf{r}-ma\hat{\mathbf{z}})d\mathbf{r}=\dfrac{1}{M}\displaystyle\sum\limits_{m=0}^{M-1}\displaystyle\int  x_m^2 p_1(\mathbf{r}_m)d\mathbf{r}_m = \eta^2\\
     \langle x \rangle & = \dfrac{1}{M}\displaystyle\sum\limits_{m=0}^{M-1}x\displaystyle\int  p_1(\mathbf{r}-ma\hat{\mathbf{z}})d\mathbf{r}=\dfrac{1}{M}\displaystyle\sum\limits_{m=0}^{M-1}\displaystyle\int  x_m p_1(\mathbf{r}_m)d\mathbf{r}_m = 0
\end{align}
(where $\mathbf{r}_m = \mathbf{r}-ma\hat{\mathbf{z}}$). Therefore $\mathcal{K}_{xx}=\eta^2=\mathcal{K}_{yy}$ and is invariant vs system size.  

Parallel to the lattice vector $a\hat{\mathbf{z}}$ however, we obtain:
\begin{align}
    \langle z^2 \rangle & = \dfrac{1}{M}\displaystyle\sum\limits_{m=0}^{M-1}\displaystyle\int z^2 p_1(\mathbf{r}-ma\hat{\mathbf{z}})d\mathbf{r}\notag \\
    &=\dfrac{1}{M}\displaystyle\sum\limits_{m=0}^{M-1}\displaystyle\int  \left(z_m+ma\right)^2 p_1(\mathbf{r}_m)d\mathbf{r}_m\notag \\
     &=\dfrac{1}{M}\displaystyle\sum\limits_{m=0}^{M-1}\left(\displaystyle\int  z_m^2 p_1(\mathbf{r}_m)d\mathbf{r}_m+m^2a^2\displaystyle\int p_1(\mathbf{r}_m)d\mathbf{r}_m+2am\displaystyle\int z_m p_1(\mathbf{r}_m)d\mathbf{r}_m\right)\notag \\
     &=\dfrac{1}{M}\displaystyle\sum\limits_{m=0}^{M-1}\left(\eta^2+m^2a^2\right)=\eta^2+\dfrac{(M-1)(2M-1)}{6}a^2\\
    \langle z \rangle & = \dfrac{1}{M}\displaystyle\sum\limits_{m=0}^{M-1}\displaystyle\int z p_1(\mathbf{r}-ma\hat{\mathbf{z}})d\mathbf{r}\notag\\
    &=\dfrac{1}{M}\displaystyle\sum\limits_{m=0}^{M-1}\displaystyle\int  \left(z_m+ma\right) p_1(\mathbf{r}_m)d\mathbf{r}_m\notag\\
    &=\dfrac{1}{M}\displaystyle\sum\limits_{m=0}^{M-1}\left(\displaystyle\int  z_m p_1(\mathbf{r}_m)d\mathbf{r}_m+ma\displaystyle\int  p_1(\mathbf{r}_m)d\mathbf{r}_m\right)\notag\\
    &=\dfrac{1}{M}\displaystyle\sum\limits_{m=0}^{M-1}ma=\dfrac{M-1}{2}a
\end{align}
\begin{align}
    \therefore \mathcal{K}_{zz}&=\langle z^2\rangle -\langle z\rangle ^2
    =\eta^2+\dfrac{(M^2-1)}{12}a^2
\end{align}
$\mathcal{K}_{zz}$ thus grows as $O(M^2)$ vs the system size $M$. This increase is however entirely due to geometric/structural factors (i.e. only dependent on the lattice spacing $a$ and not on the electronic contribution from $p_1(\mathbf{r})$). Thus differences between $\mathcal{K}_{zz}$ computed by various methods will be independent of $M$, and will solely be a function of the subsystem densities $\{p_1(\mathbf{r})\}$. This is not true in the interacting subsystems limit, but the analysis nonetheless reveals a significant contribution to $\bm{\mathcal{K}}$ from molecular geometry alone.

A similar analysis for 2D (square) and 3D (cubic) lattices shows that $\mathcal{K}_{ii}$ grows as $O(M)$ and $O(M^{\frac{2}{3}})$ respectively vs number of identical noninteracting subsystems $M$ (where $\hat{\mathbf{i}}$ is parallel to the lattice vectors), due to geometric factors, while the electronic contribution intrinsically arising from $p(\mathbf{r})$ remains constant. Consequently, differences in $\bm{\mathcal{K}}$ should be size-intensive in the non-interacting limit. However, relative error in  $\bm{\mathcal{K}}$ would shrink with increasing $M$ (as the reference value in the denominator would increase). We consequently only consider absolute errors of the form $\bm{\mathcal{K}}-\bm{\mathcal{K}}_{ref}$, vs a reference value $\bm{\mathcal{K}}_{ref}$.
This stands in contrast to properties like dipole moments and static polarizabilities, which are properly size-extensive and thus suitable for relative/percentage error based metrics. It is also possible to look at standard deviations (i.e. $\sqrt{\mathcal{K}_{ii}}$) instead of variances, but the geometric factors would prevent the errors from being size-intensive in that case. 

\begin{table}[htb!]
\begin{tabular}{|lll|ll|}
\hline
\multicolumn{3}{c}{NSP}           & \multicolumn{2}{c}{SP} \\ \hline
AlF        & Cl$_2$   & Mg        & AlH$_2$     & NH       \\
Ar         & ClCN     & Mg$_2$    & BH$_2$      & NH$_2$   \\
BF         & ClF      & N$_2$     & BO          & NO$_2$   \\
BH$_2$Cl   & FCN      & NH$_3$    & BS          & NP       \\
BH$_2$F    & H$_2$    & NH$_3$O   & Be          & Na       \\
BH$_3$     & H$_2$O   & NaCl      & BeH         & Na$_2$   \\
BHF$_2$    & HBO      & NaH       & C$_2$H      & NaLi     \\
BeH$_2$    & HBS      & Ne        & CF$_2$      & O$_2$    \\
C$_2$H$_2$ & HCCCl    & OCl$_2$   & $^3$CH$_2$  & O$_3$    \\
C$_2$H$_4$ & HCCF     & PH$_3$    & CH$_3$      & OF$_2$   \\
CH$_2$BH   & HCHO     & PH$_3$O   & CN          & P        \\
CH$_2$BO   & HCN      & SCl$_2$   & F$_2$       & P$_2$    \\
CH$_3$Cl   & HCl      & SF$_2$    & H           & PCl      \\
CH$_3$F    & HF       & SH$_2$    & H$_2$CN     & PF       \\
CH$_3$Li   & HNC      & SO$_2$    & HCHS        & PH       \\
CH$_4$     & He       & SiH$_3$Cl & HCP         & PH$_2$   \\
CO         & LiBH$_4$ & SiH$_3$F  & Li          & PO$_2$   \\
CO$_2$     & LiCl     & SiH$_4$   & N           & S$_2$    \\
CS         & LiF      & SiO       & NCl         & SO-trip  \\
CSO        & LiH      &           & NF          & SiH$_3$  \\
           &          &           & NF$_2$      &          \\ \hline
\end{tabular}
\caption{The 100 species in the dataset, sorted by whether they are not spin-polarized (NSP) or spin-polarized (SP).}
\label{tab:dataset}
\end{table}

\section{Dataset}
We have investigated $\bm{\mathcal{K}}$ for 100 small main-group systems (listed in Table \ref{tab:dataset}), for which it was possible to get highly accurate benchmark values with CC singles and doubles with perturbative triples (CCSD(T)\cite{raghavachari1989fifth}) at the complete basis set (CBS) limit.  The systems were chosen from the combined datasets considered in Refs \citenum{hait2018accurate} and \citenum{hait2018accuratepolar} such that $\bm{\mathcal{K}}$ was diagonal for a fixed coordinate system for all (spatial symmetry preserving) electronic structure methods. Asymmetric systems with nondiagonal $\bm{\mathcal{K}}$ were not included for simplicity. Linear molecules with a single $\pi$ electron/hole (such as OH) were also not considered, as real valued orbitals would be incapable of predicting a cylindrically symmetric orbital of $\pi$ symmetry \cite{lee2019kohn,brakestad2020static}, and would thus spuriously lead to $\mathcal{K}_{xx}\ne\mathcal{K}_{yy}$. Complex valued orbitals would thus be necessary to describe such species with proper symmetry within a DFT framework\cite{lee2019kohn}.

The complete dataset consists of \textcolor{black}{59} not spin-polarized (NSP) and 41 spin-polarized (SP) systems. The NSP vs SP classification was done on the basis of whether the stable HF solution has $\langle S^2\rangle =0$ or not. NSP species are thus unambiguously closed-shell, and are thus more likely to be `easier' for single-reference quantum chemistry methods like KS-DFT or MP2. \textcolor{black}{Indeed, the size of the (T) correction to $\bm{\mathcal{K}}$ was quite small for nearly all species (NSP or SP, as discussed later in Sec \ref{sec:fulldata}), indicating that the multireference character of the chosen systems (if any) did not strongly influence $\bm{\mathcal{K}}$ predictions and that CCSD(T) is likely to be an adequately accurate benchmark.}
Furthermore, none of these systems have pathological, delocalization driven qualitative failures\cite{Dutoi2006,Ruzsinszky2006,hait2018accurate,hait2018communication,vydrov2007tests}, making them reasonable choices for understanding the behavior of DFAs in the regime where they are expected to work well. 

The error $\epsilon_{i,m}$ in an individual component $\mathcal{K}^m_{ii}$ for a given molecule $m$ (vs a reference value $\mathcal{K}^m_{ref,ii}$) is:
\begin{align}
\epsilon_{i,m}&=    \mathcal{K}^m_{ii}-\mathcal{K}^m_{ref,ii}
\end{align}

The cumulative errors over all molecules and directions are thus:
\begin{enumerate}
    \item Root mean square error (RMSE): $\sqrt{\dfrac{1}{3N}\displaystyle\sum\limits_{m=1}^N\left(\epsilon_{x,m}^2+\epsilon_{y,m}^2+\epsilon_{z,m}^2\right)}$.
    \item Mean error (ME): $\dfrac{1}{3N}\displaystyle\sum\limits_{m=1}^N\left(\epsilon_{x,m}+\epsilon_{y,m}+\epsilon_{z,m}\right)$.
    \item Maximum absolute error (MAX): $\max\left(\abs{\epsilon_{i,m}}\right) \forall i\in \{x,y,z\}$ and $\forall m \in \mathbb{Z}^+, m \le N$.
\end{enumerate}
All quantities are reported in atomic units \textcolor{black}{(a.u.)} unless specified otherwise. 

\subsection{Computational Details}
All calculations were performed with the Q-Chem 5 package \cite{QCHEM4}, using fixed geometries obtained from Refs \citenum{hait2018accurate} and \citenum{hait2018accuratepolar} \textcolor{black}{(also provided in the supporting information)}. \textcolor{black}{We examined the performance of 47 DFAs spanning all five rungs of Jacob's ladder (as can be seen from Table \ref{tab:alldata}), ensuring reasonable representation at each level. Individual DFAs were selected based on (perceived) popularity, recency and performance over other benchmark datasets\cite{mardirossian2017thirty,goerigk2017look,hait2018accurate,hait2018accuratepolar,mardirossian2018survival}.}

$\bm{\mathcal{K}}$ was obtained from $\mathbf{Q}$ computed for this work and $\bm{\mu}$ obtained from Ref \citenum{hait2018accurate}.  $\mathbf{Q}$ for self-consistent field (SCF) approaches like HF and non double hybrid DFAs were found analytically via integrating over $\rho(\mathbf{r})$. $\mathbf{Q}$ for other methods (MP2, CC or double hybrid DFT) was found via a central, two point finite difference approach using a constant electric field gradient of $2\times 10^{-4}$ \textcolor{black}{a.u}. (similar to Ref \citenum{hait2018accurate}). Finite difference errors thus introduced appear to be quite small, as the RMS deviation between analytic and finite-difference CCSD/aug-cc-pCVTZ\cite{dunning1989gaussian,woon1995gaussian,peterson2002accurate,prascher2011gaussian} $\mathcal{K}_{ii}$ is $2.2\times 10^{-4}$ a.u.

Comparison between wave function theory and various DFAs are only meaningful at the CBS limit, due to different rates of basis set convergence. HF/aug-cc-pCV5Z $\mathcal{K}_{ii}$ were assumed to be at the CBS limit as it only has an RMS deviation of $2.5\times 10^{-4}$ \textcolor{black}{a.u} vs aug-cc-pCVQZ results (which should be sufficiently small, in light of the empirically observed exponential convergence of HF energies vs cardinal number of the basis set\cite{jensen2005estimating,karton2006comment}). Similarly, DFT/aug-pc-4\cite{jensen2001polarization,jensen2002polarization,jensen2002polarizationiii,jensen2004polarization,jensen2007polarization,jensen2010describing} values were assumed to be at the CBS limit for functionals from Rungs 1-4 of Jacob's ladder.  
The virtual orbital dependent correlation contribution $\bm{\mathcal{K}}^\textrm{corr}$ (in MP2/CC/double hybrids) was extrapolated to the CBS limit via the two point extrapolation formula\cite{halkier1999basis} ${\mathcal{K}^\textrm{corr,}_{ii}}^{\textrm{n}}={\mathcal{K}^\textrm{corr,}_{ii}}^{\textrm{CBS}}+\dfrac{A}{n^3}$ from aug-cc-pCVTZ ($n=3$) and aug-cc-pCVQZ ($n=4$) (which is adequate for dipoles\cite{hait2018accurate,halkier1999basis} and appears to also be adequate for quadrupole moments\cite{halkier1997props,zhang2016props,bokhan2016props}). 

Local exchange-correlation integrals for all DFT calculations were computed over
a radial grid with 99 points and an angular Lebedev grid with 590 points for all atoms. Non-local correlation was evaluated on an SG-1 grid\cite{gill1993standard}. Unrestricted (U) orbitals were employed for all CC (except Be, where UHF breaks spatial symmetry) and non-double hybrid DFT calculations. MP2 is known to yield non N-representable densities when spin-symmetry breaks\cite{Kurlancheek2009}, leading to poor dipole\cite{hait2018accurate} and polarizability\cite{hait2018accuratepolar} predictions. Consequently, MP2 with both restricted (R) and unrestricted orbitals were carried out, and the restricted variant found to yield a significantly smaller RMSE. Double hybrid calculations were subsequently carried out with only R orbitals. Stability analysis was performed in the absence of fields to ensure the orbitals correspond to a minimum, and the resulting orbitals were employed as initial guesses for finite field calculations (if any). The frozen-core approximation was not employed in any calculation.

\section{Results and discussions}

\begin{table}[htb!]
\scriptsize
\begin{tabular}{@{}lllllllllllllll@{}}
\hline
Method	&	Class		&	\multicolumn{3}{c}{RMSE}	&	ME	&	MAX	&	&	Method	&	Class	&	\multicolumn{3}{c}{RMSE}	&	ME	&	MAX	\\	\hline								
	&	&			Full	&	NSP	&	SP	&	&	&	&	&	&						Full	&	NSP	&	SP	&	&	\\		
CCSD     & WFT    & 0.003  & 0.002  & 0.003  & -0.001 & 0.022 &  & SCAN0\cite{scan0}      & Rung 4 & 0.008  & 0.004  & 0.011  & 0.000  & 0.047 \\
MP2      & WFT    & 0.022  & 0.004  & 0.035  & 0.001  & 0.312 &  & B97-2\cite{b97-2}      & Rung 4 & 0.009  & 0.004  & 0.013  & 0.001  & 0.047 \\
RMP2     & WFT    & 0.010  & 0.004  & 0.015  & 0.002  & 0.074 &  & TPSSh\cite{tpssh}      & Rung 4 & 0.009  & 0.004  & 0.013  & 0.002  & 0.053 \\
HF       & WFT    & 0.027  & 0.015  & 0.039  & 0.007  & 0.301 &  & PBE0\cite{pbe0}       & Rung 4 & 0.009  & 0.005  & 0.014  & 0.003  & 0.052 \\
RHF      & WFT    & 0.020  & 0.015  & 0.026  & 0.006  & 0.091 &  & B97\cite{b97}        & Rung 4 & 0.010  & 0.006  & 0.013  & 0.005  & 0.049 \\
         &        &        &        &        &        &       &  & HSE-HJS\cite{hsehjs}    & Rung 4 & 0.010  & 0.005  & 0.014  & 0.003  & 0.051 \\
SPW92\cite{Slater,PW92}    & Rung 1 & 0.022  & 0.017  & 0.028  & 0.011  & 0.129 &  & SOGGA11-X\cite{sogga11x}  & Rung 4 & 0.010  & 0.005  & 0.015  & 0.004  & 0.057 \\
Slater\cite{Slater}   & Rung 1 & 0.047  & 0.038  & 0.057  & 0.034  & 0.275 &  & LRC-$\omega$PBEh\cite{lrcwpbeh}  & Rung 4 & 0.011  & 0.005  & 0.016  & 0.002  & 0.061 \\
         &        &        &        &        &        &       &  & PW6B95\cite{pw6b95}     & Rung 4 & 0.011  & 0.006  & 0.016  & 0.003  & 0.084 \\
BPBE\cite{b88,PBE}     & Rung 2 & 0.013  & 0.008  & 0.017  & 0.004  & 0.062 &  & $\omega$M05-D\cite{wM05D}     & Rung 4 & 0.011  & 0.007  & 0.015  & 0.005  & 0.064 \\
mPW91\cite{mpw91}    & Rung 2 & 0.014  & 0.010  & 0.019  & 0.006  & 0.069 &  & M06\cite{m06}        & Rung 4 & 0.011  & 0.007  & 0.016  & 0.001  & 0.058 \\
B97-D3\cite{b97d}   & Rung 2 & 0.016  & 0.010  & 0.022  & 0.007  & 0.103 &  & $\omega$B97X-V\cite{wb97xv}    & Rung 4 & 0.012  & 0.008  & 0.016  & 0.006  & 0.066 \\
PBE\cite{PBE}      & Rung 2 & 0.016  & 0.012  & 0.021  & 0.008  & 0.079 &  & $\omega$B97X-D\cite{wB97XD}    & Rung 4 & 0.012  & 0.006  & 0.017  & 0.003  & 0.064 \\
N12\cite{n12}      & Rung 2 & 0.020  & 0.013  & 0.028  & 0.006  & 0.120 &  & M06-2X\cite{m06}     & Rung 4 & 0.013  & 0.006  & 0.019  & 0.003  & 0.104 \\
BLYP\cite{b88,lyp}     & Rung 2 & 0.021  & 0.016  & 0.026  & 0.012  & 0.112 &  & HFLYP\cite{lyp}      & Rung 4 & 0.014  & 0.013  & 0.015  & -0.004 & 0.051 \\
SOGGA11\cite{sogga11}  & Rung 2 & 0.048  & 0.035  & 0.062  & 0.024  & 0.500 &  & M11\cite{m11}        & Rung 4 & 0.014  & 0.009  & 0.019  & 0.005  & 0.083 \\
         &        &        &        &        &        &       &  & B3LYP\cite{b3lyp}      & Rung 4 & 0.014  & 0.009  & 0.019  & 0.007  & 0.068 \\
SCAN\cite{SCAN}     & Rung 3 & 0.009  & 0.005  & 0.012  & 0.002  & 0.052 &  & CAM-B3LYP\cite{camb3lyp}  & Rung 4 & 0.014  & 0.010  & 0.019  & 0.007  & 0.070 \\
MS2\cite{ms2}      & Rung 3 & 0.010  & 0.004  & 0.014  & 0.000  & 0.072 &  & M08-HX\cite{m08so}     & Rung 4 & 0.015  & 0.009  & 0.020  & 0.009  & 0.079 \\
TPSS\cite{tpss}     & Rung 3 & 0.010  & 0.006  & 0.015  & 0.003  & 0.065 &  & $\omega$B97M-V\cite{wB97MV}    & Rung 4 & 0.022  & 0.013  & 0.030  & 0.013  & 0.127 \\
mBEEF\cite{wellendorff2014mbeef}    & Rung 3 & 0.011  & 0.005  & 0.017  & -0.001 & 0.074 &  & MN15\cite{MN15}       & Rung 4 & 0.035  & 0.013  & 0.052  & 0.013  & 0.231 \\
M06-L\cite{m06l}    & Rung 3 & 0.014  & 0.010  & 0.018  & -0.006 & 0.068 &  &            &        &        &        &        &        &       \\
revM06-L\cite{revm06l} & Rung 3 & 0.015  & 0.013  & 0.018  & 0.002  & 0.091 &  & DSD-PBEPBE\cite{kozuch2013spin}     & Rung 5 & 0.005  & 0.003  & 0.007  & 0.002  & 0.032 \\
B97M-V\cite{b97mv}   & Rung 3 & 0.026  & 0.013  & 0.037  & 0.011  & 0.157 &  & XYGJ-OS\cite{zhang2011fast}     & Rung 5 & 0.006  & 0.003  & 0.008  & 0.002  & 0.028 \\
M11-L\cite{m11l}    & Rung 3 & 0.038  & 0.023  & 0.052  & 0.005  & 0.332 &  & PTPSS\cite{goerigk2010efficient}      & Rung 5 & 0.006  & 0.003  & 0.009  & 0.002  & 0.032 \\
MN15-L\cite{mn15l}   & Rung 3 & 0.040  & 0.024  & 0.055  & 0.022  & 0.278 &  & XYG3\cite{zhang2009doubly}       & Rung 5 & 0.006  & 0.003  & 0.009  & 0.002  & 0.032 \\
         &        &        &        &        &        &       &  & $\omega$B97M(2)\cite{mardirossian2018survival}   & Rung 5 & 0.008  & 0.005  & 0.010  & 0.003  & 0.037 \\
         &        &        &        &        &        &       &  & B2GPPLYP\cite{karton2008highly}   & Rung 5 & 0.008  & 0.005  & 0.011  & 0.004  & 0.042 \\
         &        &        &        &        &        &       &  & $\omega$B97X-2(TQZ)\cite{chai2009long} & Rung 5 & 0.008  & 0.006  & 0.010  & 0.005  & 0.044 \\
         &        &        &        &        &        &       &  & B2PLYP\cite{grimme2006semiempirical}     & Rung 5 & 0.010  & 0.007  & 0.013  & 0.005  & 0.047\\
         \hline							
\end{tabular}
\caption{Errors in $\bm{\mathcal{K}}$ (\textcolor{black}{in a.u}) predicted by various DFAs for the species in the dataset. \textcolor{black}{Positive ME values indicate less compact (i.e. more diffuse) densities, relative to the benchmark.}}
\label{tab:alldata}
\end{table}

\subsection{Full dataset}\label{sec:fulldata}

The errors in DFA predictions for the full dataset are reported in Table \ref{tab:alldata}, along with errors for the wave function methods CCSD, MP2 and HF. Considering the wave function theories (WFTs) first, we see that CCSD has the lowest RMSE of 0.003 of all the methods considered, and gives fairly similar performance across the NSP and SP subsets. MP2 gives good performance for the NSP dataset, but N-representability failures\cite{Kurlancheek2009} lead to significantly worse results over the SP subset. Use of R orbitals ameliorates this considerably, with RMP2 having a quite improved RMSE of 0.010 over the full dataset. RMP2 is nonetheless still somewhat challenged by open-shell systems (especially NF and NCl) where the artificial enforcement spin-symmetry appears to be suboptimal. \textcolor{black}{We further note that MP3 is also expected to have the same N-representability failures as MP2 and is thus unlikely to lead to significant improvements over MP2. This in fact has been observed for $\mu$, although alternative orbital choices could lead to better MP performance.\cite{rettig2020third}}

HF performs quite poorly due to lack of correlation, having an RMSE of 0.027 (which can be considered as a ceiling for reasonable DFA performance). Based on the mean error (ME), HF has a tendency to make the variance too large (i.e. HF densities are too diffuse).  The SP species are much more challenging than NSP, and RHF actually reduces RMSE substantially to 0.020. However, this is almost solely on the account of two challenging alkali metal containing species (Na$_2$ and NaLi) and the RMSE for HF and RHF are quite similar upon their exclusion. Nonetheless, this serves as a warning that spin-symmetry breaking might compromise property predictions, despite being the natural approach for improving the energy. In particular, spin-symmetry breaking in diatomics like Na$_2$ or NaLi results in densities similar to two independent atoms vs a bonded molecule, leading to larger widths than the restricted solution. 

Coming to the DFAs, we observe that rung 1 local spin-density approximations (LSDA) functionals fare considerably worse than CCSD or RMP2. Bare Slater \textcolor{black}{(LSDA)} exchange has a rather large RMSE of 0.047, while inclusion of correlation reduces RMSE substantially to 0.022 for SPW92. The SP subset is significantly more challenging in both cases, with an RMSE that is almost double the corresponding NSP value. The minimally parameterized nature of the LSDA functionals mean that the SPW92 RMSE of 0.022 can also be considered as a reasonable reference for judging DFAs against. In fact, the HF, RHF and SPW92 RMSEs collectively suggest that DFAs with RMSE larger than 0.02 are not really accurate for predicting $\bm{\mathcal{K}}$. 

Moving up Jacob's ladder to rung 2 generalized gradient approximations (GGAs), we observe that all but one of the functionals investigated improve upon SPW92. SOGGA11 is the exception, being essentially as bad as bare Slater exchange. On the other hand, BPBE and mPW91 have substantially lower RMSEs, indicating considerable improvement over rung 1. However, the SP RMSE continues to be nearly twice the NSP one for all GGAs. 

Rung 3 meta-GGAs (mGGAs) further improve upon predictions, with the best mGGA (SCAN) predicting a quite low RMSE of 0.009, while MS2 and TPSS are also fairly reasonable. It is worth noting that all three were developed with a heavy emphasis on non-empirical constraints and were specifically fit to the H atom (one of the most challenging species in the dataset, as discussed later). In contrast, modern, mGGAs developed principally by fitting to empirical benchmark data appear to fare worse, with MN15-L, M11-L and B97M-V being particularly disappointing as they fare worse than SPW92. Nonetheless, it is worth noting that the modern, empirically fitted mBEEF functional yields a respectable performance. 

Hybrid functionals on rung 4 of Jacob's ladder do not significantly improve upon mGGAs. The best-performer is SCAN0, which only marginally improves upon the parent SCAN functional. Similarly, TPSSh only slightly improves upon TPSS. This general behavior stands in contrast to the case of dipole moments, where hybrid functionals strongly improve upon lower rungs\cite{hait2018accurate}. Some other decent performers are B97-2 (which was partially fitted to densities) PBE0, B97, HSE-HJS and SOGGA11-X which also happen to be hybrid GGAs. PW6B95 is the best-performing hybrid mGGA that was not already based on an existing Rung 3 functional with good performance. 
Interestingly, HFLYP improves significantly upon HF, mostly by virtue of dramatically reducing SP errors (in large part because it does not break spin-symmetry for NaLi or Na$_2$). SP errors are roughly 2-3 times the NSP errors for nearly all other hybrid functionals. It is also worth noting that that modern $\omega$B97M-V and MN15 functionals fare particularly poorly, yielding performance worse than SPW92. On the other hand, several empirically fitted functionals like M06 and $\omega$B97X-V yield quite respectable performance (albeit significantly worse than SCAN0). 

Rung 5 double hybrid functionals reduce error significantly vs hybrids, with DSD-PBEPBE approaching CCSD levels of RMSE for NSP species (although SP species are much more challenging). Even B2PLYP (the poorest performer, and coincidentally, the oldest) performs similarly to a good hybrid functional like PBE0.  Most rung 5 functionals also improve upon RMP2 (especially for the SP subset), indicating perceptible benefit from the local exchange-correlation contributions. It is also worth noting that the recently developed $\omega$B97M(2) functional yields mediocre performance, despite being one of the best-performers for energy predictions\cite{mardirossian2018survival} (although it represents an enormous improvement over the parent $\omega$B97M-V functional). 

The positive ME values in Table \ref{tab:alldata} also indicates that most DFAs predict slightly less compact densities (especially LSDA). This is perhaps not too surprising in light of delocalization error present in the studied functionals\cite{hait2018delocalization}.  However, the connection between density compactness and delocalization error is not always straightforward, as shown later. Indeed, it can be seen that the local M06-L  and mBEEF functionals systematically predict too small $\bm{\mathcal{K}}$, which is contrary to expectations based on delocalization error alone. HFLYP however has a negative ME, consistent with overlocalization of density that is expected from 100\% HF exchange. On the other hand, HF systematically overestimates $\bm{\mathcal{K}}$, which seems puzzling at a first glace as HF should overlocalize if anything\cite{li2017piecewise,hait2018delocalization}. However, lack of correlation in HF leads to artificial symmetry breaking where densities become more `atom-like' than `bond-like', leading to spurious overestimation of $\bm{\mathcal{K}}$ for NaLi and Na$_2$. In fact, lack of correlation hinders electrons from occupying the same region of space and could lead to wider `spread' in the electron density. 

Comparison to the results obtained earlier for dipole moments\cite{hait2018accurate} indicates some similarities and differences. Ascending Jacob's ladder leads to improved predictions in both cases, and several modern, empirically designed functionals are found to perform relatively poorly. In fact, some of the functionals that were found to be amongst the top performers for their rung in the dipole study (SCAN, DSD-PBEPBE, PBE0, SOGGA11-X etc.) continue to perform well. The major differences are that hybrid functionals do not significantly improve upon mGGAs and some functionals that are good for $\bm{\mathcal{K}}$ predictions do not fare as well for dipoles. A clear example of this is SCAN0 faring worse at predicting $\bm{\mu}$ than $\omega$B97M-V, despite the former being the best hybrid for predicting $\bm{\mathcal{K}}$ (while the latter is amongst the worst). Nonetheless, the ability of several functionals to predict both $\bm{\mathcal{K}}$ and $\bm{\mu}$ with low error is encouraging, as that indicates reasonable performance for problems involving external electric fields.

It is also worthwhile to identify what species in the dataset are most challenging for DFAs. Ordering the molecules in the dataset by the first quartile of RMS error (over cartesian axes) across different DFAs reveals that there is a break in the distribution after the first 10 species. These difficult cases are: Be, H, Li, NaLi, Na$_2$, BeH, LiH, H$_2$, BeH$_2$ and BF in descending order of difficulty. Several of these cases are also known to be challenging for dipole moment\cite{hait2018accurate} and static polarizability\cite{hait2018accuratepolar} predictions as well. NaLi in particular was already known to be a very challenging case for many DFAs, and it is unsurprising that the analogous Na$_2$ is also challenging. \textcolor{black}{Both of these alkali metal dimers feature long bonds ($\sim$ 3 {\AA}) and are SP at equilibrium geometry, suggesting some multireference character (analogous to the isovalent case of stretched H$_2$). In addition, Na$_2$ and NaLi are outliers with respect to RMS deviation of CCSD $\mathcal{K}$ vs the CCSD(T) benchmark, having $\sim$ 0.01 deviation (while the next largest deviation is about half of that, for CH$_3$Li). Near-exact calculations with the adaptive sampling configuration interaction method\cite{tubman2016deterministic,tubman2020modern} (with the small cc-pVDZ basis\cite{prascher2011gaussian}) however indicate that the species are not particularly multireference (a single determinant has $\sim$ 90\% weight in the total wave function) and CCSD(T) is sufficiently accurate for $\mathcal{K}$ (as shown in the supporting information).
Furthermore}, the three atoms represent the most challenging cases overall, with Li having a first quartile RMSE $\sim$ 35\% larger than NaLi. 

\subsection{The case of challenging atoms}
\begin{figure}[htb!]
\includegraphics[width=\textwidth]{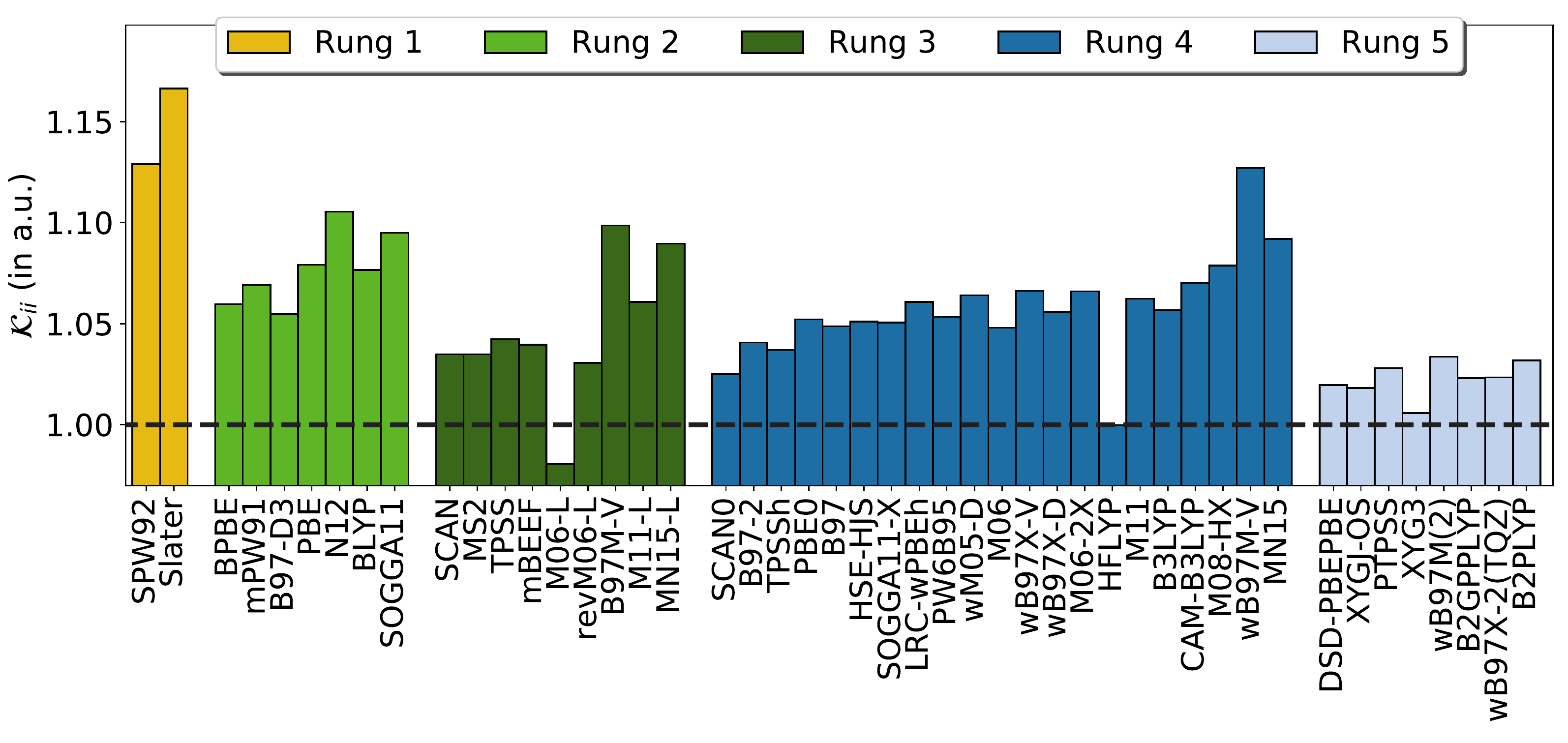}
\caption{$\mathcal{K}_{ii}$ predicted for the H atom, by various DFAs (ordered by their position in Table \ref{tab:alldata}). The dashed line is the analytic value.}
\label{fig:H_error}
\end{figure}
It is thus worthwhile to examine these atoms in greater detail for better insight into functional behavior. 
H has only one electron and thus all DFA errors are definitionally self-interaction driven. It is thus interesting to observe that the error in $\bm{\mathcal{K}}$ does not appear to correlate well with the delocalization error in many cases, as can be seen from Fig \ref{fig:H_error}. Perhaps the most clear example of this is a comparison between the local M06-L functional and the range separated hybrid $\omega$B97M-V, with the former surprisingly predicting a \textit{smaller} H atom than exact quantum mechanics, while the latter overestimates the size to essentially the same extent as SPW92 (contrary to the behavior seen for delocalization error\cite{hait2018delocalization}). Similarly, M06-2X predicts a larger H atom than M06 despite having twice the HF exchange (54\% vs 27\%).
It is thus apparent that the local exchange-correlation components of several modern density functionals lead to significant errors for the size of the H atom, which are counterintuitive from the perspective of delocalization error (or fraction of HF exchange present). 

\textcolor{black}{On the other hand, if we only consider functionals that use the same local exchange-correlation components hybridized with varying amounts of HF exchange (such as PBE/PBE0), we find that error decreases with increase in the HF exchange contribution. This is on account of the functional becoming strictly more HF like.}
It is also worth noting that the double hybrid functionals appear to have much lower error than hybrids, potentially due to relatively smaller contributions from local exchange-correlation.  While cancellation of errors in larger systems is likely to make this less of an issue (as can be seen from other species in the dataset), future density functional development would likely benefit from including the size and polarizability\cite{hait2018accuratepolar} of the H atom as soft constraints.

\begin{figure}[htb!]
\includegraphics[width=\textwidth]{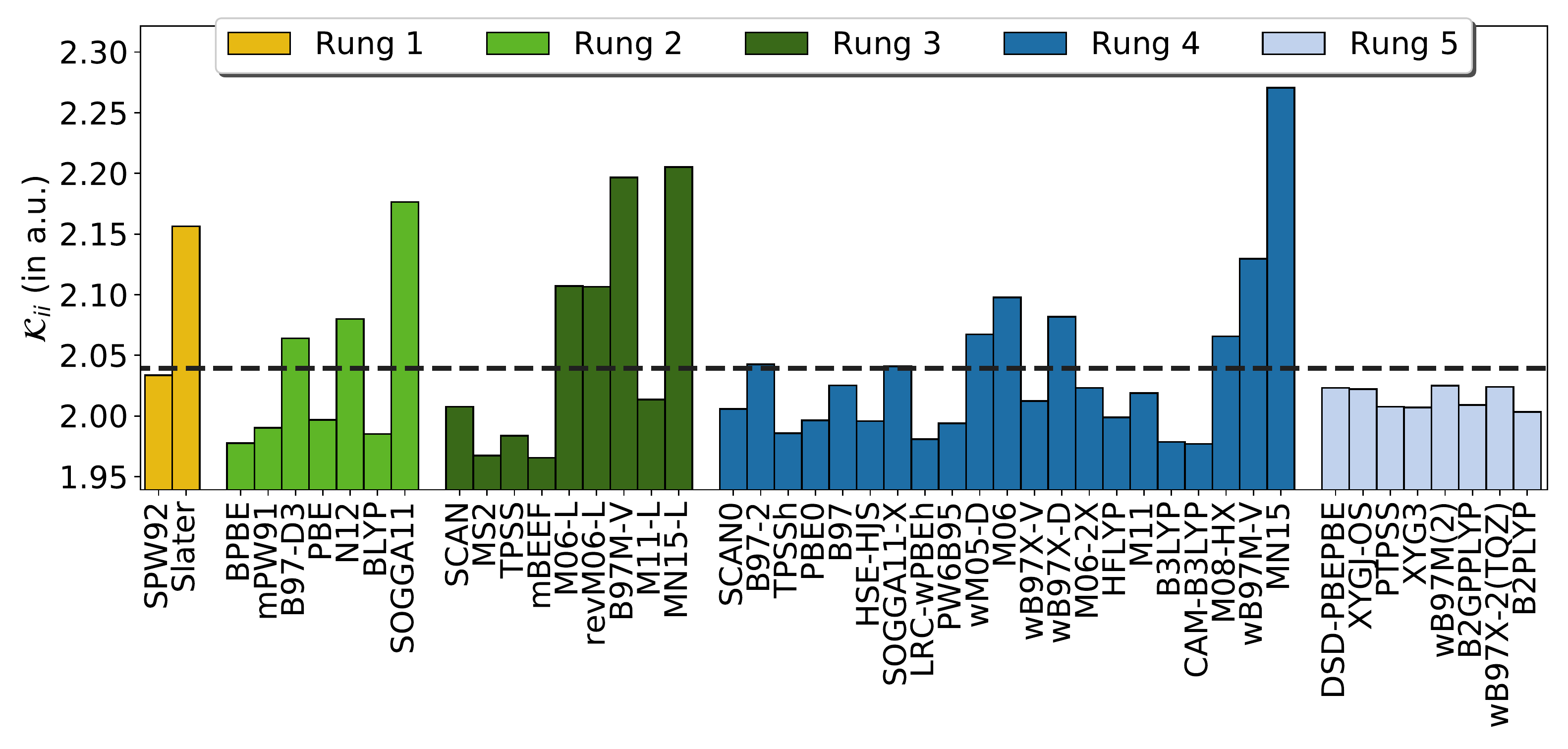}
\caption{$\mathcal{K}_{ii}$ predicted for the Li atom, by various DFAs. The dashed line is the benchmark (CCSD(T)/CBS) value and the HF value is 2.070.}
\label{fig:Li_error}
\end{figure}

Li superficially resembles H if the core electrons are neglected. However, Fig \ref{fig:Li_error} shows that the error $\bm{\mathcal{K}}$ predictions are quite different. Most functionals \textit{underestimate} the size of the diffuse alkali metal atom, predicting a more compact density than the benchmark. \textcolor{black}{Delocalization errors are normally expected to lead to less compact densities due to self-interaction. This can indeed be seen from the overly repulsive mean-field electron-electron repulsion potentials predicted by many DFAs\cite{gaiduk2012self,gritsenko2016errors}. The Li atom is also not multireference, and thus the reason behind the size underestimation by many functionals is not entirely clear. }
HFLYP not being an outlier with respect to overlocalization indicates that delocalization is unlikely to be a major factor (and bare HF in fact overestimates the size).
Nonetheless, most of the modern functionals that fare poorly for H (B97M-V, MN15-L, MN15 and $\omega$B97M-V) continue to fare poorly for Li and overestimate the size, revealing considerable scope for improvement. The size of the Li atom is thus another reasonable choice as a soft constraint for future functional fitting. 

\begin{figure}[htb!]
\includegraphics[width=\textwidth]{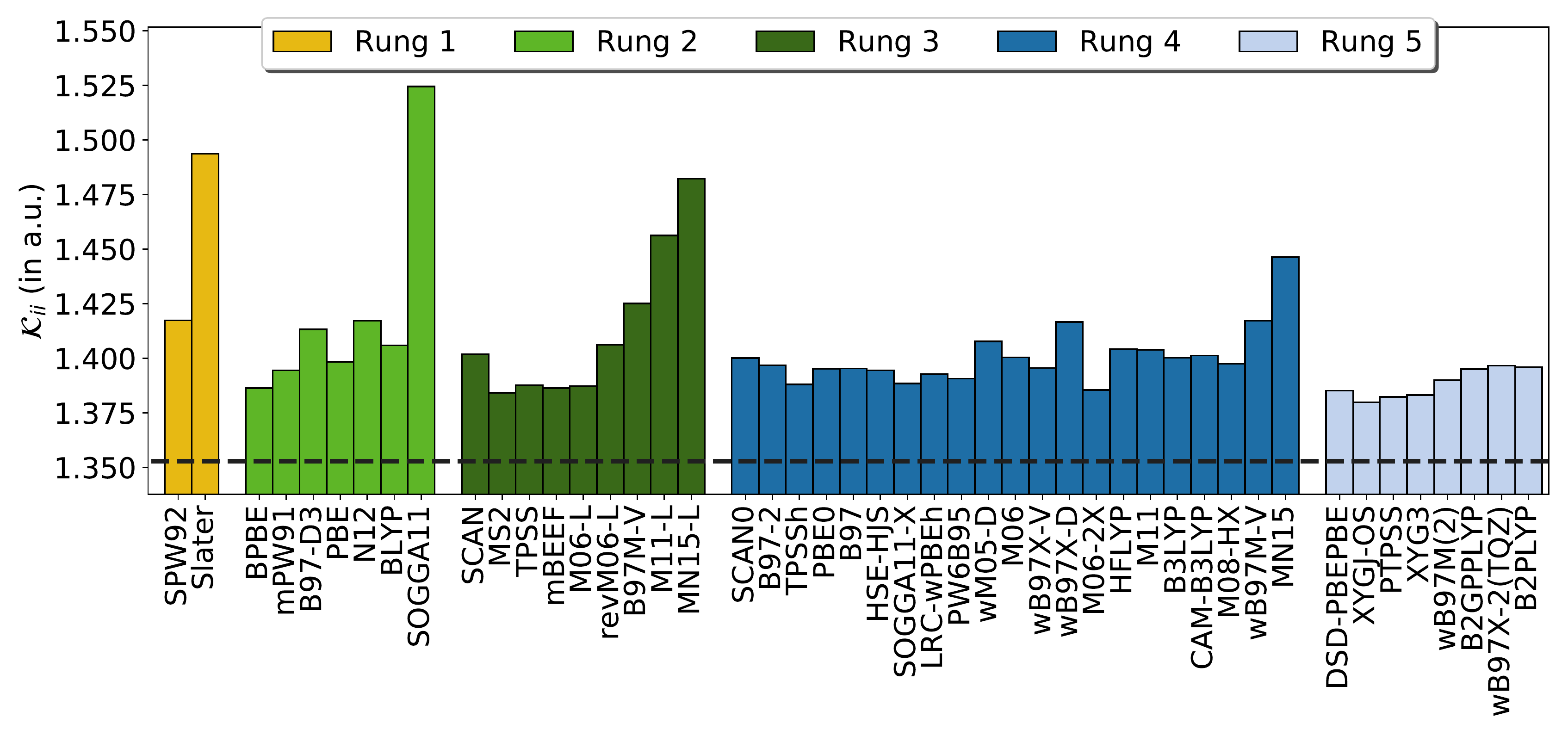}
\caption{$\mathcal{K}_{ii}$ predicted for the Be atom, by various DFAs. The dashed line is the benchmark (CCSD(T)/CBS) value and the HF value is 1.444.}
\label{fig:Be_error}
\end{figure}

In contrast to the preceding two cases, Fig \ref{fig:Be_error} shows that most methods overestimate the size of the Be atom by a similar amount, with some poor performing outliers. In fact, the best mGGA and hybrid have $\sim 0.001$ variation in $\mathcal{K}_{ii}$ (while the best double hybrid is better by $\sim 0.005$). 
This relatively uniform performance is likely a consequence of the challenges faced by DFAs in accounting for the multireference character of the Be atom (it has 0.37 effectively unpaired electrons\cite{head2003characterizing}). It is nonetheless worth noting that the worst performers are recently parameterized modern functionals. It is not entirely clear whether the Be atom should feature in datasets used for functional training because of the partial multireference character of this problem, but it should definitely be employed in test sets to gauge fit quality. 

\begin{figure}[htb!]
\begin{minipage}{0.48\textwidth}
\includegraphics[width=\textwidth]{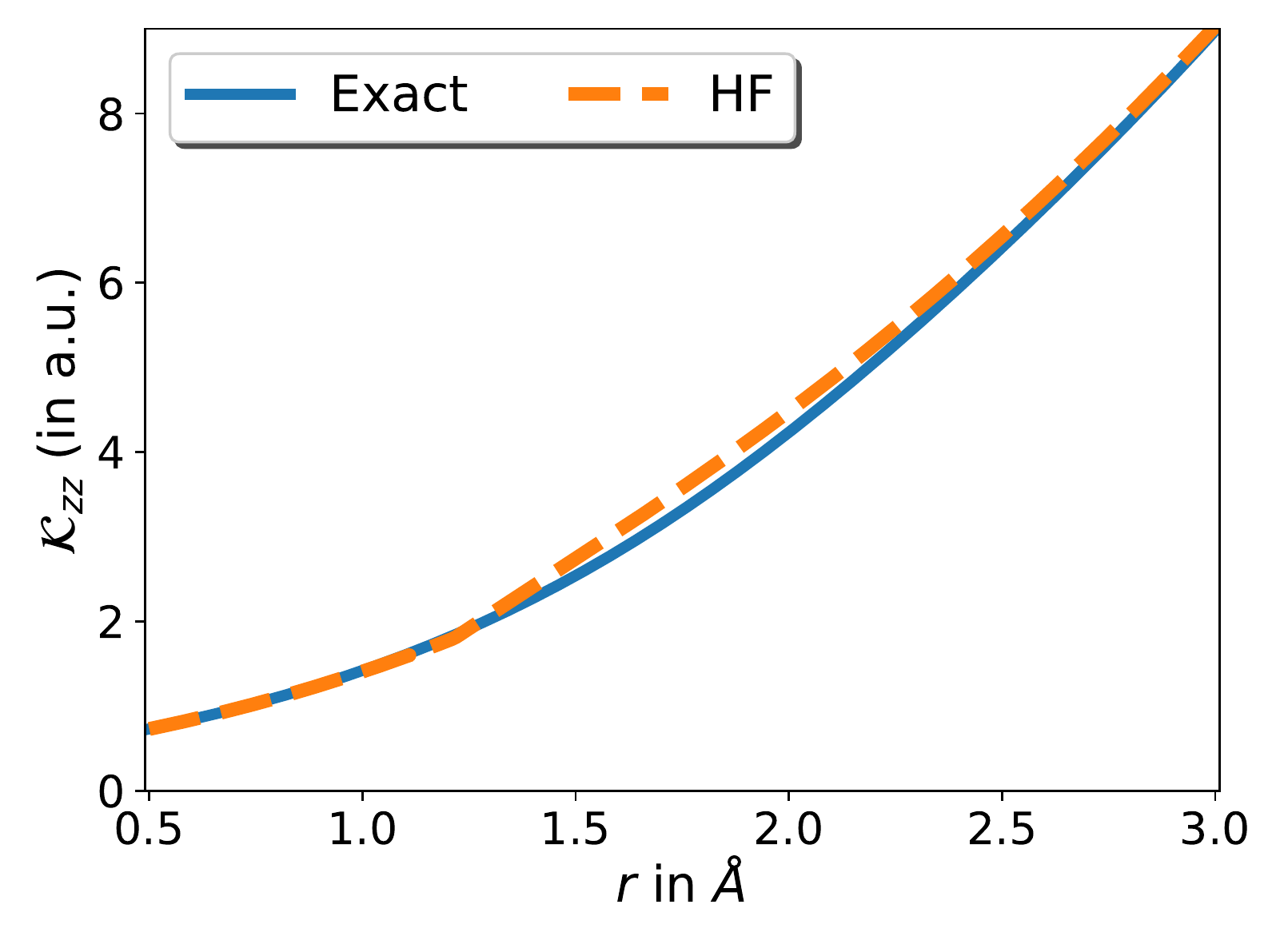}
\subcaption{Exact $\mathcal{K}_{zz}$ values, along with HF.}
\label{fig:h2_actual}
\end{minipage}
\begin{minipage}{0.48\textwidth}
\includegraphics[width=\textwidth]{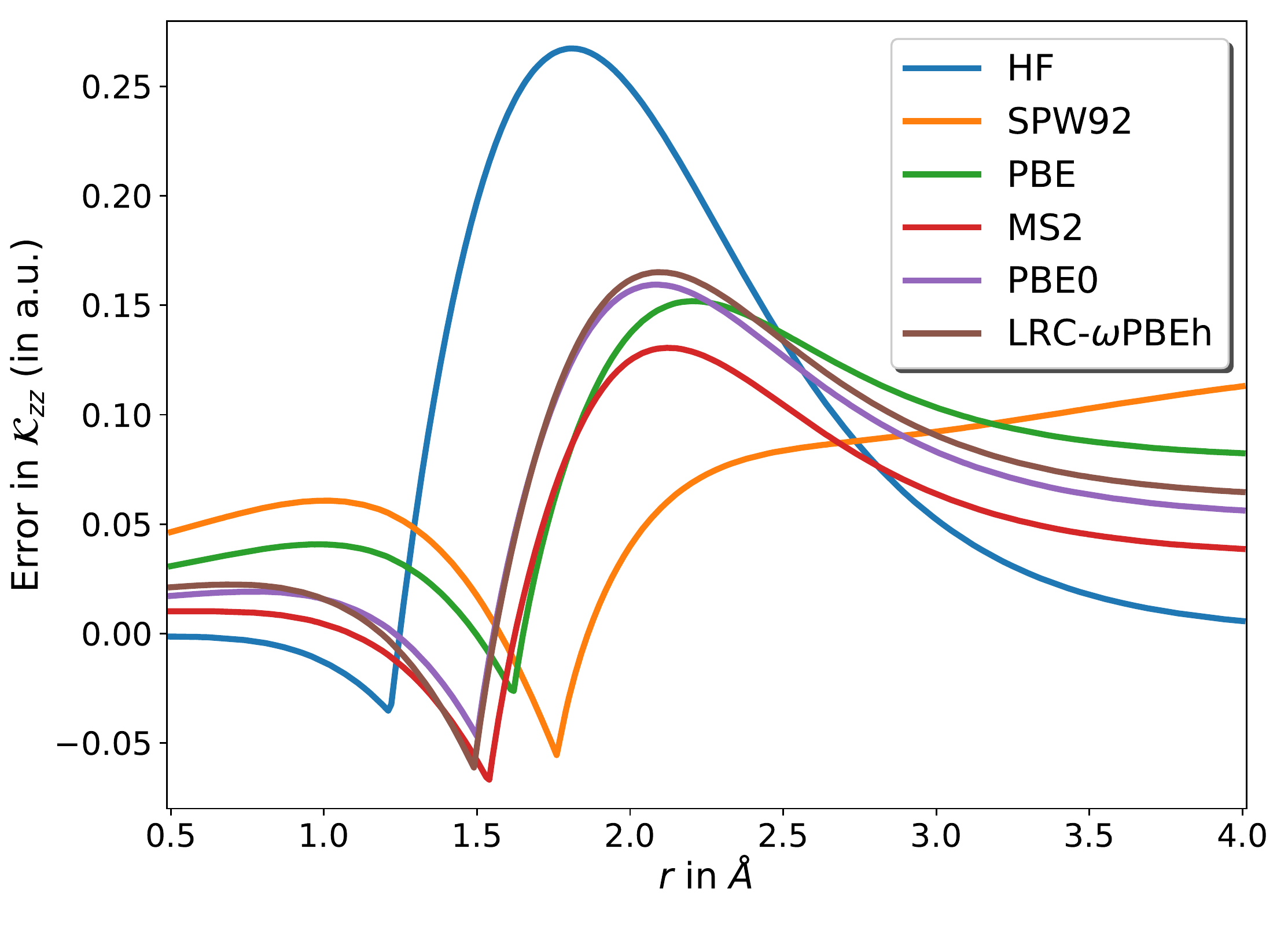}
\subcaption{Errors in $\mathcal{K}_{zz}$ predicted by various DFAs.}
\label{fig:h2_error}
\end{minipage}
\caption{Behavior of $\mathcal{K}_{zz}$  for stretched H$_2$.}
\end{figure}
 
\section{Stretched H$_2$}
The behavior of DFAs in predicting $\bm{\mathcal{K}}$ in non-equilibrium configurations is potentially interesting. We only investigate the case of H$_2$, to avoid qualitative failures associated with delocalization error for polar bonds\cite{Dutoi2006,Ruzsinszky2006,hait2018accurate,hait2018communication} for general bond dissociations. We furthermore restrict ourselves to a few well-behaved functional at Rung 4 and below, to avoid incomplete spin-polarization based problems\cite{hait2019beyond} . Double hybrid functionals are not considered due to N-representability problems in UMP2,\cite{Kurlancheek2009} and divergence of RMP2 with bond stretching (indicating that they are quite unsuitable for bond dissociation problems). 
Fig \ref{fig:h2_actual} shows that $\mathcal{K}_{zz}$ grows quadratically with the internuclear distance $r$, and Fig \ref{fig:h2_error} thus only plots the error in $\mathcal{K}_{zz}$ for clarity. All of the single-reference methods encounter a derivative discontinuity at the Coulson-Fischer (CF) point\cite{coulson1949xxxiv} (which is the onset of spin-polarization) due to the orbital stability matrix being singular at that point. The behavior right beyond this point is the most interesting, as the two electrons are fairly strongly correlated in this regime. 
The DFAs other than SPW92 yield remarkably similar behavior in this region, systematically overestimating $\mathcal{K}_{zz}$ somewhat (and thus predicting more `atom' like densities as opposed to a still partially bonded molecule). HF exhibits the same qualitative behavior, but has much larger errors due to complete absence of correlation (indeed, the systematic overestimation past the CF point can be clearly observed in Fig \ref{fig:h2_error} as well). Similar overestimation in $\bm{\mathcal{K}}$ is seen for the Be atom, which also has partial multireference character. Consistent systematic overestimation of $\bm{\mathcal{K}}$ by several well-behaved functionals with varying levels of delocalization error could in fact potentially be a signature of strong correlation, as the electrons are forced to be less compact in the absence of explicit (i.e. post mean-field) correlation that allows them to occupy the same region of space. 

\section{Discussion and Conclusion}
The objective of this work was to characterize the electron density of small molecules using a scalar metric that goes beyond the first moment of the electron density (i.e. dipole moments). Since the molecular quadrupole moment is origin-dependent for molecules with non-zero charge or dipole, we elected instead to characterize the $3 \times 3$ matrix of second cumulants, or spatial variances, of the electron density, $\bm{\mathcal{K}}$. The eigenvalues of $\bm{\mathcal{K}}$ are thus a measure of the square of the characteristic extent of the molecular density along each principal axis.

We produced a benchmark dataset, \textcolor{black}{Var213}, that contains \textcolor{black}{213} benchmark values of $\bm{\mathcal{K}}$, from 100 small molecules, evaluated with CCSD(T)/CBS correlation based on extrapolation towards the complete basis set limit from aug-cc-pCVTZ and aug-cc-pCVQZ calculations, combined with HF/aug-cc-pCV5Z. These reference values were then used to assess $\bm{\mathcal{K}}$ predictions from 47 density functional approximations (DFAs). Broadly, the molecules studied here are relatively straightforward in terms of their electronic structure. In the language of ref. \citenum{mardirossian2017thirty}, this dataset should be considered ``easy'' for DFAs, rather than ``difficult'' due to the absence of strong correlation or delocalization effects.

The results show that it is possible to obtain quite reasonable $\bm{\mathcal{K}}$ values in many cases, even with modern density functionals. However, non-empirical functionals, especially at the mGGA level via SCAN, and at the hybrid level via SCAN0, seem to fare better than more empirically parameterized models. This constitutes useful independent validation of the quality of electron densities from these functionals. Double hybrid rung 5 functionals yield the best overall performance: indeed all double hybrids tested match or exceed the performance of SCAN, and only the early B2PLYP double hybrid fails to outperform SCAN0. 

Interestingly, the performance of the modern MN15-L, MN15, B97M-V and $\omega$B97M-V functionals over the studied dataset is disappointing, suggesting that the vastness of the meta-GGA space\cite{b97mv} has offered these functionals the opportunity to predict reasonable energies from fairly poor densities. There is an interesting lesson to be drawn from this outcome, particularly for the minimally parametrized meta-GGA-based functionals such as B97M-V and $\omega$B97M-V whose 12-16 parameter form was inferred from the use of large nunbers of chemically relevant energy differences. Specifically, such forms are evidently not fully constrained by the data used to generate them. Independent data such as $\bm{\mathcal{K}}$ may indeed be very useful in attempting to develop improved data-driven functional designs.

Overall, our $\bm{\mathcal{K}}$ benchmark tests show that the performance of the best-performing functional strictly improves on ascending Jacob's ladder, indicating that extra complexity has the potential to improve behavior. 
Use of this dataset and other density based information (and especially the H, Li and Be atoms) for future functional development could thus provide sufficient soft constraints to yield improved $\bm{\mathcal{K}}$ values, and by implication electron densities as well. The results obtained here may thereby contribute to offering a route to better computationally tractable approximations to the exact functional.

\section*{Acknowledgment} 
	This research was initially supported by the Director, Office of Science, Office of Basic Energy Sciences, of the U.S. Department of Energy under Contract No. DE-AC02-05CH11231. Completion of the project was based upon work performed by the Liquid Sunlight Alliance, a DOE Energy Innovation Hub, supported by the U.S. Department of Energy, Office of Science, Office of Basic Energy Sciences, under Award Number DE-SC0021266. Y.H.L. was funded via the UC Berkeley College of Chemistry summer research program.
\section*{Supporting Information} 
	\noindent Geometries of species studied (zip).
	\noindent Computed values and analysis (xlxs). 
	
\section*{Data Availability}
The data that supports the findings of this study are available within the article and its supplementary material.
\bibliography{references}
\end{document}